\documentclass[conference,letter]{IEEEtran}
\usepackage{subcaption}     
\usepackage{makecell}
\usepackage{graphicx}
\usepackage{latexsym}
\usepackage{amsxtra} 
\usepackage{amsmath}
\usepackage{amsfonts}
\usepackage{marvosym}
\usepackage{scalefnt}
\usepackage{color,tabularx}
\usepackage{makecell}
\usepackage{booktabs,multicol,multirow}
\usepackage{algorithm}
\usepackage{comment}
\usepackage{url}
\usepackage{soul}
\usepackage{cite}
\usepackage{xcolor}
\usepackage[noend]{algpseudocode}
\usepackage{hyperref}

\IEEEoverridecommandlockouts \IEEEpubid{\makebox[\columnwidth]{979-8-3503-5934-3/24/\$31.00~\copyright{}2024 IEEE \hfill} \hspace{\columnsep}\makebox[\columnwidth]{ }}

\usepackage{fancyhdr}
\fancypagestyle{firstpage}{
  \fancyhf{}
  \fancyhead[C]{ 	To appear at the 2024 27th International Symposium on Design \& Diagnostics of Electronic Circuits \& Systems (DDECS).}
  \fancyfoot[L]{979-8-3503-5934-3/24/\$31.00~\copyright{}2024 IEEE }
}

\begin{document}\bstctlcite{IEEEexample:BSTcontrol}
\title{
Exploring Quantization and Mapping Synergy in Hardware-Aware Deep Neural Network Accelerators
}

\author{
 \IEEEauthorblockN{Jan Klhufek, Miroslav Safar, Vojtech Mrazek, Zdenek Vasicek, Lukas Sekanina}
 \IEEEauthorblockA{\scalefont{0.95}{Brno University of Technology, Faculty of Information Technology, Brno, Czech Republic} \\
 Email: iklhufek@fit.vutbr.cz, xsafar23@stud.fit.vutbr.cz, \{mrazek, vasicek, sekanina\}@fit.vutbr.cz}
}

\maketitle

\thispagestyle{firstpage}

\begin{abstract}

Energy efficiency and memory footprint of a convolutional neural network (CNN) implemented on a CNN inference accelerator depend on many factors, including a weight quantization strategy (i.e., data types and bit-widths) and mapping (i.e., placement and scheduling of DNN elementary operations on hardware units of the accelerator). We show that enabling rich mixed quantization schemes during the implementation can open a previously hidden space of mappings that utilize the hardware resources more effectively. CNNs utilizing quantized weights and activations and suitable mappings can significantly improve trade-offs among the accuracy, energy, and memory requirements compared to less carefully optimized CNN implementations. To find, analyze, and exploit these mappings, we: (i) extend a general-purpose state-of-the-art mapping tool (Timeloop) to support mixed quantization, which is not currently available; (ii) propose an efficient multi-objective optimization algorithm to find the most suitable bit-widths and mapping for each DNN layer executed on the accelerator; and (iii) conduct a detailed experimental evaluation to validate the proposed method. On two CNNs (MobileNetV1 and MobileNetV2) and two accelerators (Eyeriss and Simba) we show that for a given quality metric (such as the accuracy on ImageNet), energy savings are up to 37\% without any accuracy drop.

\end{abstract}

\section{Introduction}

Highly optimized hardware accelerators and memory subsystems have been developed to meet the challenging energy, performance, and memory requirements imposed on the fast and efficient processing of \emph{convolutional neural networks} (CNNs)~\cite{sze:pieee17}. While CNNs are typically trained on GPUs that utilize floating point (FP) data representation, the most energy-efficient CNN hardware (inference) accelerators employ spatial architecture and advanced mixed-precision quantization schemes operating with a few bits~\cite{ArmeniakosZSH23}. We will primarily deal with \emph{weight quantization}, which can significantly reduce memory size and transfers in addition to performance and energy efficiency gains. Furthermore, we will also quantize the activations, which mainly affect the smaller internal scratchpad memory. As quantization almost always negatively impacts the CNN’s error, various automated quantization methods enabling suitable trade-offs between the error and hardware parameters (such as inference energy, latency, and weight memory size) have been proposed~\cite{HAQ:2019}.

\begin{figure}[t]
    \vspace*{-0.3em}
    \includegraphics[width=\columnwidth]{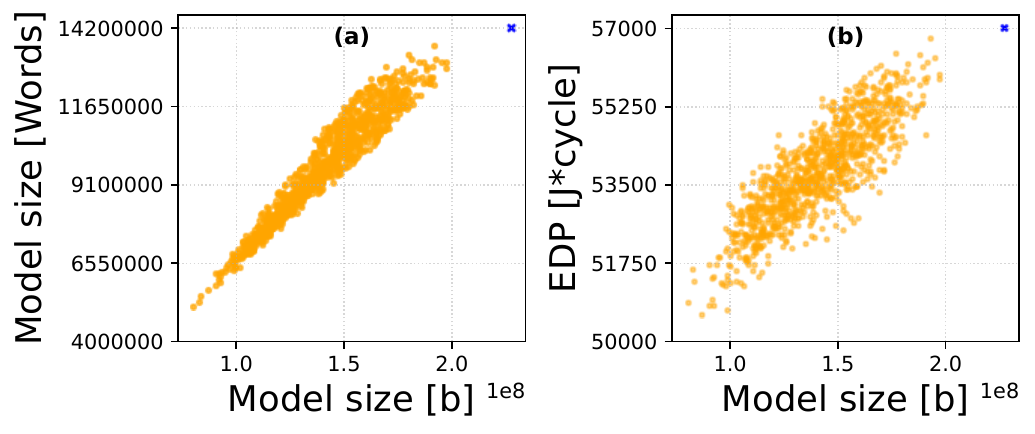}
    \vspace*{-2em}
    \caption{Correlation between Model Size (the number of bits for weights) and: \textbf{(a)} Memory Word Count after bit-packing \textbf{(b)} and evaluated Energy-Delay Product on Eyeriss for $1000$ unique configurations of MobileNetV1. The blue point represents the original uniform setting.}
    \vspace*{-1.9em}
    \label{fig:naive_vs_measured_comparison}
\end{figure}

In this paper, we focus on one aspect of CNN quantization not addressed by the literature: performance gains that can be obtained when the synergy of a suitable \emph{quantization} and \emph{mapping} (i.e., placement and scheduling of CNN elementary operations on hardware units) is exploited in CNN hardware accelerators. Consider a  mixed-precision quantization method guided by na\"ive memory footprint estimation, e.g., a layer-wise optimization of weight bit-widths minimizing the total number of bits in weight memory, i.e., the model size. Fig.~\ref{fig:naive_vs_measured_comparison}(a) shows that when the na\"ive estimation is applied in the MobileNetV1 CNN, the correlation with the memory word count (after bit-packing) is far from perfect (calculated for 1000 randomly generated quantized versions of MobileNetV1). Moreover, only a weak correlation is visible for the Energy-Delay-Product (EDP) evaluated for one inference on the Eyeriss accelerator (Fig.~\ref{fig:naive_vs_measured_comparison}(b)). These weak correlations exist because the accelerator’s architecture, mapping, and memory subsystem, which significantly influence hardware parameters, are not considered during quantization. 

\IEEEpubidadjcol

The use of mixed-precision layer-wise quantization schemes is motivated by the fact that different layers have different error resilience characteristics and behave differently on the hardware~\cite{ArmeniakosZSH23}. This observation can be exploited by various optimization and approximation strategies. 
However, the mixed-precision quantization is not supported by state-of-the-art configurable general-purpose models of hardware accelerators utilized to estimate the hardware parameters of a given CNN implementation before its deployment. It turns out that the automated quantization methods only consider either the CNN model (not the hardware) \cite{Ristretto} or exist for a particular hardware accelerator \cite{HAQ:2019}, i.e., they are not usable across a reasonable set of CNN accelerators. This is partly because determining hardware parameters using real hardware accelerators is time-consuming.

To easily model, analyze, and investigate the impact of various mixed-precision quantization schemes (particularly those based on computing the most suitable bit-width for weights and activations for each layer) on the hardware parameters of CNN hardware accelerators, we extended Timeloop to support different mixed-precision quantization schemes. Timeloop~\cite{TimeLoop:2019} is a popular tool searching for the most suitable mapping of the CNN model on the accelerator resources, in which a rich set of accelerator architectures, configurations, and mapping algorithms can be utilized. When coupled with Accelergy~\cite{Accelergy:iccad:2019}, an early-stage energy and execution time estimation tool, detailed hardware characteristics can be obtained quickly. Then, we show that enabling rich mixed quantization schemes during the implementation of a CNN can open a previously hidden space of mappings that utilize the hardware resources more effectively than uniformly quantized layers accompanied by standard mappings. CNNs utilizing quantized weights and activations (both input and output) and suitable mappings can significantly improve trade-offs among the accuracy, energy, and memory requirements compared to less carefully optimized CNN implementations. We propose to determine the most suitable mixed-precision quantization schemes with a multi-objective genetic algorithm NSGA-II~\cite{deb:2002} (with extended Timeloop in the loop) that simultaneously minimizes the weight memory size (reflecting the accelerator’s memory subsystems), inference energy, and CNN error. The optimal mapping is obtained using Timeloop. To stress the importance of mixed-precision quantization, we intentionally keep the accelerator specification unchanged for all experiments.

In this paper, we make the following key contributions.
\begin{enumerate}
\item To investigate the impact of mixed-precision quantization schemes on CNN accelerator’s performance and energy consumption, we extend Timeloop to support mixed-precision quantization. We show that adopting the mixed precision quantization can increase the number of valid (CNN model-to-hardware) mappings by one order of magnitude, enabling thus many optimization strategies unreachable with a common approach.
The extended version of Timeloop is available as open-source code at: {\textcolor{blue}{\url{https://github.com/ehw-fit/timeloop-with-quantization}}}
\item To analyze the impact of a particular CNN quantization on the CNN error, we connect extended Timeloop with the PyTorch framework. The proposed extension can be alternatively connected with the TensorFlow framework. 
\item To find a Pareto front containing CNNs that show high-quality tradeoffs between the error and hardware parameters, we propose to adopt NSGA-II with Timeloop in the loop. Timeloop coupled with Accelergy is much faster than accelerator simulation or in-hardware measurement so that many candidate solutions can be quickly generated and evaluated. The method is parallelized and executed much faster than other approaches such as~\cite{HAQ:2019} and allows the exploration of various accelerator architectures. 
\item On two CNNs (MobileNetV1 and MobileNetV2) and two CNN accelerators (Eyeriss and Simba), we show that for a given quality metric (such as classification accuracy on a subset of ImageNet), energy savings are up to 37\% without any accuracy drop. 
\end{enumerate}

\section{Related Work}
\label{sec:soa}

\emph{CNN hardware accelerators} with spatial architecture employ an array of locally communicating processing elements (PE). Each of them implements a multiply-and-accumulate circuit, a small local memory (registers), and a controller. PEs are usually organized as pipelined systolic arrays optimized for fast execution of CNN operations. The term \emph{mapping} refers to the dataflow strategy (i.e., computation order and parallelism strategy) coupled with the tiling strategy (selecting the size of input data with respect to available hardware resources such as buffers and PEs). A particular mapping of a given CNN on hardware resources and the execution of this mapping are implemented by the accelerator controller. The accelerator can use a common dataflow strategy (e.g., input stationary, row stationary)~\cite{sze:pieee17} or highly optimized dataflow determined using methods such as GAMMA (Genetic Algorithm-based Mapper for ML Accelerators)~\cite{GAMMA:2020}. In addition to optimizing the dataflow, common approaches to reach energy-efficient processing and low memory footprint are weight compression, CNN pruning, arithmetic operation approximation, and quantization~\cite{ArmeniakosZSH23}. A survey of quantization methods for efficient neural network inference was presented in~\cite{Gholami:21}. The quantization can be uniform, non-uniform, symmetric, asymmetric, static, and dynamic. It can be applied layerwise, groupwise, channelwise, and sub-channelwise. Quantization-aware training simulating quantization noise during F32 training by inserting fake quantization to the model~\cite{abs-1712-05877} and post-training quantization of the pre-trained F32 model are two major quantization strategies. The mix-precision quantization in which several quantization precisions are applied together, can further improve the common uniform quantization.
If the automated CNN model design is performed together with the hardware optimization, we speak about hardware-aware neural architecture search~\cite{APQ:2020,NAAS:DAC21}. A candidate CNN implementation is evaluated in terms of hardware parameters (such as energy and latency) using either simulation (which is usually slow), analytical modeling (often not reliable), estimation tools (such as Timeloop~\cite{TimeLoop:2019}, Accelergy~\cite{Accelergy:iccad:2019}, BitFusion~\cite{BitFusion:2018}, MAESTRO~\cite{Maestro:21}, DNN chip predictor~\cite{DNNChipPredictor:2020}), surrogate ML models (fast but less reliable), or measurements on real hardware~\cite{Sze:how:evaluate:2020}.

Unfortunately, current tools developed for hardware parameter estimation of CNN inference accelerators do not support mixed-precision quantization, which is crucial for lowering energy consumption and memory utilization.

\section{Proposed framework}
\label{sec:proposed}
\begin{figure}[t]
    \centering
    \includegraphics[width=.8\columnwidth]{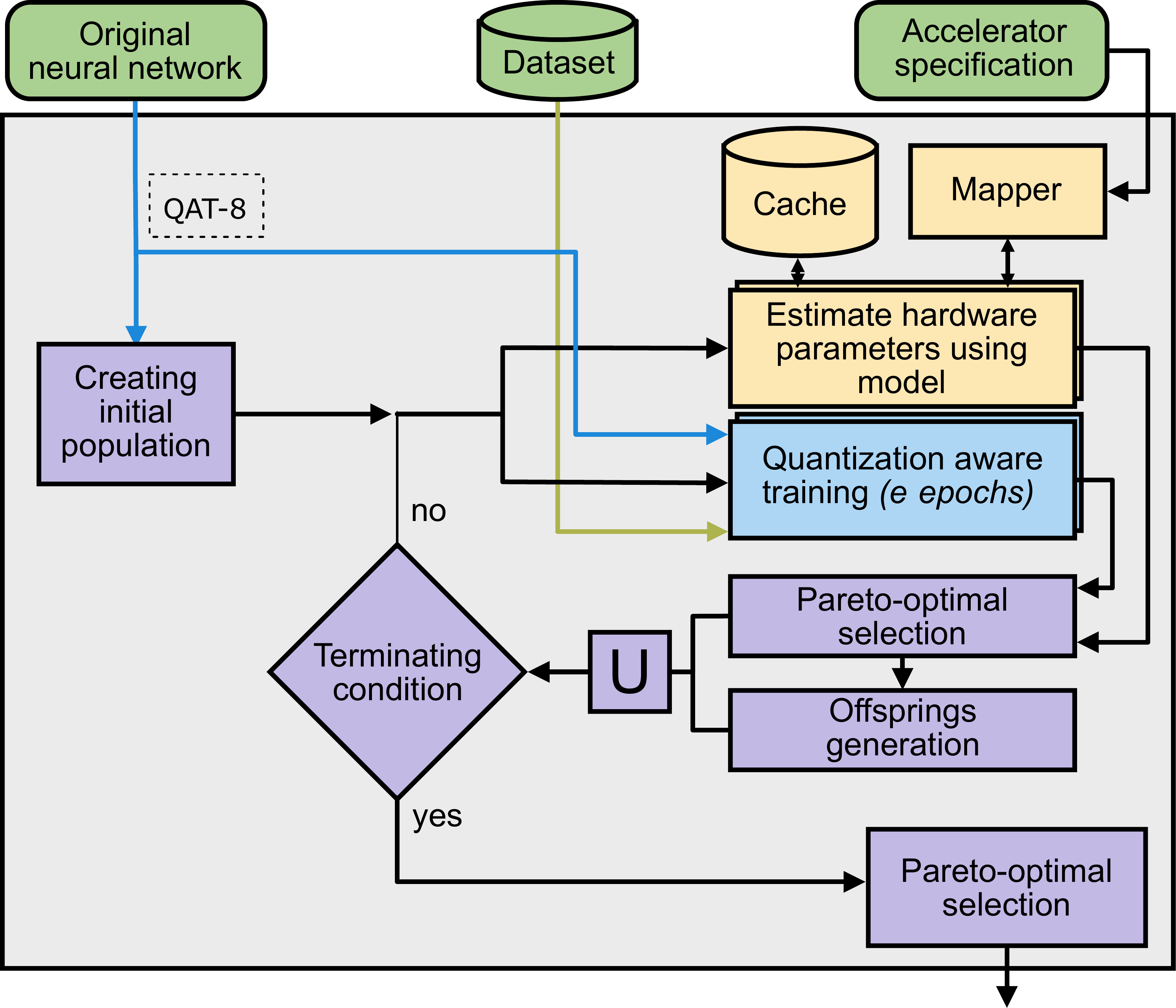}
    \caption{Overall scheme of the proposed quantization and accelerator-mapping optimization. Search engine (purple components), Training engine (blue), Mapping engine (yellow).}
    \label{fig:overall}
    \vspace{-1.5em}
\end{figure}

The proposed framework consists of three components (a) a mapping engine, (b) a training engine, and (c) a search engine. 
The overall scheme is shown in Fig.~\ref{fig:overall}.
The search engine is responsible for generating configurations based on which a quantized version of the original CNN is obtained. 
The goal of the search engine is to minimize CNN error and hardware resources.
Given a hardware accelerator, the mapping engine determines the best mapping for the obtained quantized CNN. The mapper consumes a configuration file describing the computational problem, parameters of the target HW accelerator, and constraints.
It outputs the best possible CNN execution plan
and additional information, such as the number of memory accesses, latency, and consumed energy that are needed to run the search engine. 
The training engine is used to perform retraining after quantization to recover from a quality loss 
caused by introducing the quantized layers. 
To obtain the best configurations, the search process is iterated for a predefined number of iterations.
Both mapping and training engines must support mixed-precision quantization.

\subsection{Mapping engine}
 The mapping engine is based on Timeloop~\cite{TimeLoop:2019} and Accelergy~\cite{Accelergy:iccad:2019}, which we have extended to support mixed-precision mapping.
Timeloop internally uses a mapper that finds the best way to schedule operations and stage data on the specified architecture. Various non-deterministic as well as deterministic heuristics are implemented in the original software to explore the design space effectively. 

Our extension consists of the following modifications. First, the specification is extended to permit the specification of bit-widths associated with the data tensors of individual workloads. Second, the mapping algorithm is modified to support so-called bit-packing~\cite{Bitpacking:2018}.
This technique simply allows as many data elements (i.e. weights or activations) to be stored in a single memory word as the bit-width of the word allows.
When properly implemented, this technique not only allows for minimizing the amount of memory used for storing weights and activations but also allows for maximizing the memory bus utilization because multiple weights could be loaded in parallel. 
This results in fewer memory accesses and lower energy consumption for data transfer compared to a na\"ive approach, where one memory word is occupied by one weight/activation. 
In addition to that, we modified a checker which checks for mapping violations.

Timeloop is used to characterize the workload on a per-layer basis. That is, scheduling is computed for each layer of the quantized network independently. To fully exploit the impact of quantized activations on memory utilization, the bit-width of both inputs and outputs are considered. The output bit-width is determined by the bit-width of the following layer's inputs, while constant $8$ bits are set for the last layer's outputs. The total energy is determined as a sum of the energies required to compute every workload. The same is valid also for total latency.
Because we are using the mapping engine in a loop, we have taken advantage of the per-layer approach, and we have implemented a caching mechanism. Once a layer workload has been evaluated, the results are stored in a cache. Subsequently, the cached results can be read and reused when trying to find the best plan for the same workload, eliminating the need for re-evaluation. This mechanism helps to accelerate substantially the design space exploration because the candidate configurations typically contain many similar parts.


\subsection{Training engine}
The training engine is built on top of PyTorch framework and the principles of Quantization Aware Training (QAT). We chose the PyTorch framework because of its scalability, broad support on various computing platforms, and good support of quantization.


The quantization is based on a per-tensor asymmetric approach, which is natively supported in PyTorch. According to the tested configuration, the bit-widths of the weights and the activations of the quantizable layers are set. PyTorch can only handle 8-bit quantization; lower bit-widths are implemented using specialized so-called observer modules that modify the allowed range of values. Then, the fusing of layers to include batch normalization is applied, and finally, fake-quantization observers are added. 
The modified model can then be trained and validated.

When evaluating accuracy by running QAT, we typically start with a pre-trained network at FP32 accuracy. We can quantize this network for a given bit-width and perform training for $e$ epochs during the optimization loop. Since the QAT is the most time-consuming part of the optimization and needs to be performed as fast as possible, it is feasible to pre-quantize the input model to some higher number of bits (e.g., 8) and only perform fine-tuning for a smaller number of $e$ epochs in the loop.

\subsection{Search engine}
In the mapping engine, only the memory path is considered in this study to highlight the impact of mixed-precision weight quantization on energy and memory usage. The computational MAC units remain untouched.

The accelerator configuration corresponding to the recipe for obtaining the quantized network from the original network is modeled using a linear string of tuples of integers. For MobileNetV1, for example, the string consists of 56 integers. 
Each tuple corresponds to a single layer and determines the bit-width of the inputs and weights of the associated layer. The bit-width of the outputs is determined by the bit-width of the inputs of the subsequent layer. 

We use a multi-objective genetic algorithm (NSGA-II)~\cite{deb:2002} to obtain 
quantized CNNs showing desired properties.
The algorithm maintains a population of $|P|$ candidate configurations. 
The search starts from a population consisting of configurations corresponding with uniformly quantized CNNs.
The candidate configurations are then iteratively optimized with respect to the accuracy of quantized CNN and EDP required to perform one inference. 
For each candidate solution, a corresponding PyTorch model and YAML configuration are created and utilized in the training and mapping engine. 
Then, the energy and the accuracy of the quantized CNN are evaluated. 

The new population of candidate configurations is created from the current parent population $P$ by the so-called uniform crossover followed by the mutation operator. The number of offspring individuals is $|Q|$.
The uniform crossover randomly picks two configurations from $P$ and produces a single offspring configuration 
where each integer is chosen with equal probability from one of the two parents. 

Then, with probability $p_{mutAcc}$ one randomly selected layer of the offspring is set to the initial quantization 8/8, and with the probability $p_{mut}$, one randomly selected integer is replaced by a randomly generated but valid value.
At the end of the optimization process, when a terminating condition is met (typically, the maximum number of allowed iterations is exceeded), 
the error of quantized CNNs is evaluated using the complete training set. 
Solutions whose parameters are dominated by at least one other solution are filtered out.

\section{Experimental Setup}

Two pre-trained CNN models are considered: MobileNetV1 and MobileNetV2.
For the evaluation, we classified images from the ImageNet dataset. To speed up the evaluation, we have chosen a subset consisting of 100 classes.
Both models were trained using FP32 representation for 150 epochs (with top-1 accuracies 77.26\% and 77.86\% respectively) and then optionally quantized to 8 bits for 50 epochs.
These quantized networks (FP32 and QAT-8 variants), as well as the dataset, serve as input (see Original neural network and Dataset, respectively, in~Figure~\ref{fig:overall}) to our tool.
The proposed method is evaluated independently for two HW accelerators supported by Timeloop/Accelergy: Eyeriss and Simba.
Eyeriss consists of 168 16-bit PEs, Simba employs 256 16-bit PEs.
The memory word size is 16.
The characterization is done for 45nm technology.
The accelerators are provided to our tool in form of a text specification (see ~Figure~\ref{fig:overall}).

The number of epochs for QAT is $e = 10$ for FP32 model and $e=5$ for QAT-8 model since it has already been adapted to the quantization. For the QAT-8 model we also considered $e=\{10,20\}$ number of epochs.
The experiments were executed on HPC nodes equipped with two AMD 7763 64-core CPUs and 8 NVIDIA A100 GPUs with 40 GB HBM2 RAM. The search is terminated after 48 hours.

The NSGA-II setup was determined after several test runs: the weights and activations in each layer of CNN are allowed to be quantized to 2 -- 8 bits; $|P|=32$; $|Q|$ is chosen from $\{8,16,32\}$; $p_{mut}=10\%$; $p_{mutAcc} = 5\%$.
It means that up to 32 Pareto-optimal solutions can be generated at the end.
Timeloop mapper is configured to use random search with termination condition set to finding $2000$ valid mappings per workload.

\section{Results}
\label{sec:results}

\begin{figure*}[t]
    \centering
    \includegraphics[width=1\textwidth]{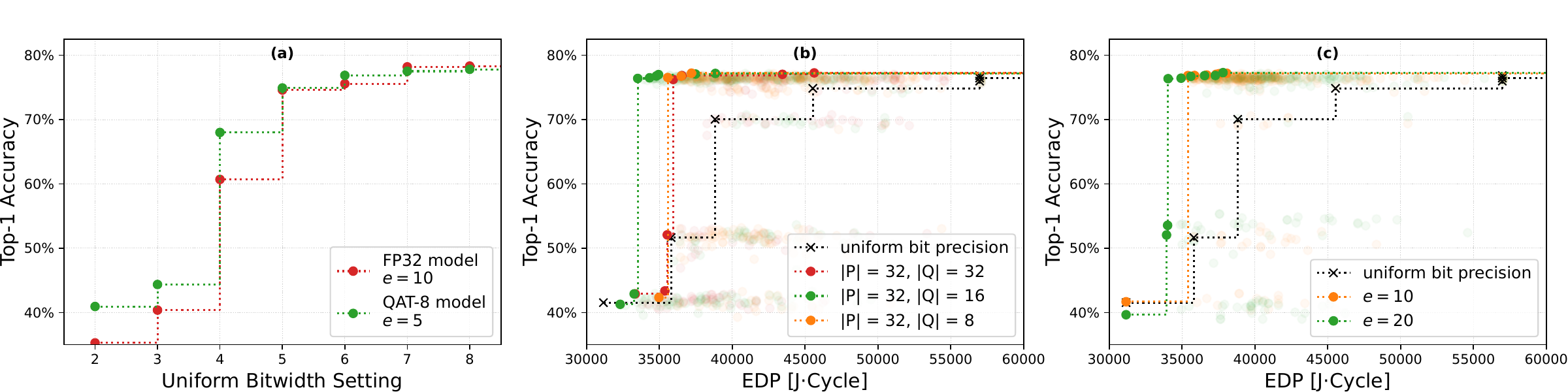}    
    \caption{Final Pareto fronts for (a) different Original neural network used for QAT fine-tuning, (b) different settings for the number of offsprings $|Q|$, and (c) different number of epochs $e$ during QAT.}
    \label{fig:final}
    \vspace{-1.5em}
\end{figure*}

\begin{table}[b]
    \centering
    \caption{The total number of exhaustively evaluated valid mappings for the second convolutional layer of the MobileNet models on two HW accelerators for various quantization settings. EDP of the least energy demanding mapping is given in J$\cdot$Cycles.  
    }
    \label{tab:mappings_eyeriss_simba}
    \resizebox{\columnwidth}{!}{
    
    \begin{tabular}{c|c|cc|cc}
    \toprule
        \bf Network & \bf Bit-width Settings & \multicolumn{2}{c|}{\textbf{Eyeriss}} & \multicolumn{2}{c}{\textbf{Simba}}\\
        \bf Layer & $q_a,q_w,q_o$ & Mappings & min. EDP  & Mappings & min. EDP  \\
    \midrule

\multirow{6}{*}{\makecell{\bf{MobileNet}\\ conv. layer \#2}}            
                                      & $16$, $16$, $16$ & $11,778$ & $638$ & $80,835$ & $204$ \\
                                      & $8$, $8$, $8$    & $15,021$ & $388$ & $110,032$ & $159$ \\
                                      & $8$, $4$, $8$    & $15,054$ & $372$ & $111,090$ & $159$ \\
                                      & $8$, $2$, $8$    & $15,054$ & $364$ & $112,113$ & $158$ \\
                                      & $4$, $4$, $4$    & $16,417$ & $281$ & $127,214$ & $137$ \\
                                      & $2$, $2$, $2$    & $16,877$ & $228$ & $133,568$ & $126$ \\
    \bottomrule
    \end{tabular}
    }
\end{table}

\subsection{Quantization-aware mapping}
The purpose of the first experiment is to quantify the impact of the modifications introduced in Timeloop on a) the size of the space of possible valid mappings and b) energy savings.
Table~\ref{tab:mappings_eyeriss_simba} shows the number of exhaustively enumerated valid mappings for various quantization settings for one convolutional layer.
To conduct this analysis, we selected the second convolutional layer (a depthwise convolutional layer) present in both analyzed variants of MobileNet.
Symbols $q_a, q_o$ denote the activation and partial sum bit-width; $q_w$ is the weight bit-width.

In the default setting, both accelerators employ 16-bit operands. For this setting, the search space consists of 11,778 and 80,835 valid mappings for Eyeriss and Simba. 
are considered, the number of valid mappings increases to 15,021 and 110,032, respectively. 
Keeping $q_a$ and $q_o$ at 8 bits and further reducing $q_w$ will slightly increase the number of mappings. This effect is visible, especially for the Simba accelerator, mainly due to the fact that Eyeriss employs the row stationary dataflow. On the other hand, decreasing both the activations and the weights bit-width results in a more noticeable increase in the number of valid mappings. This is because the execution of some mappings is bottle-necked by memories associated with storing the activations.
The results show that the proposed modification of Timeloop allows for significantly extending the amount of valid mappings.

\begin{figure}[ht]
    \vspace{-0.5em}
    \centering
    \includegraphics[width=\columnwidth]{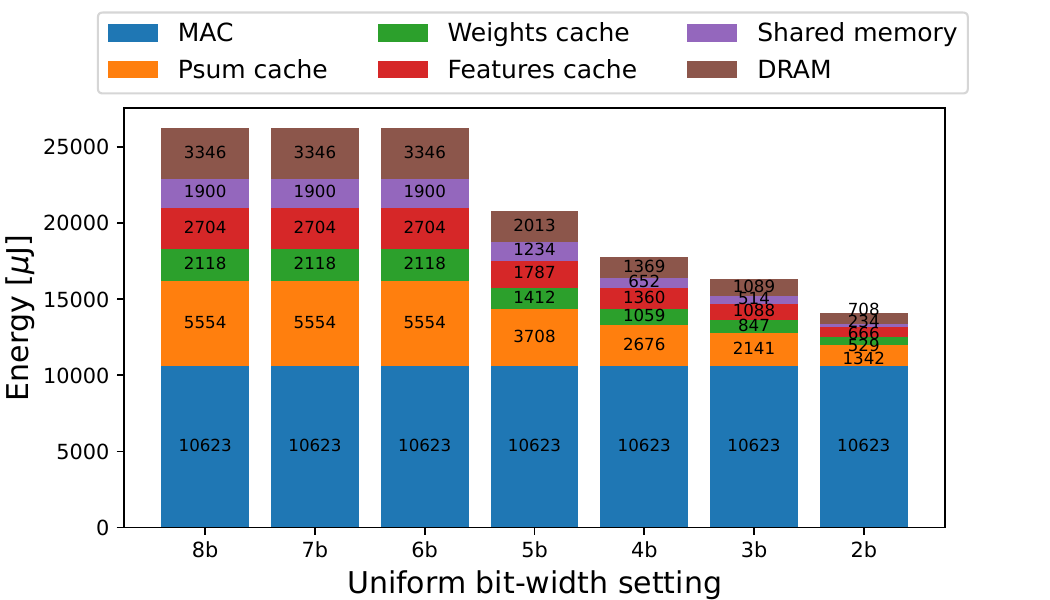}
    \vspace{-1.5em}
    \caption{Energy breakdown of various versions of quantized MobileNetV1 on Eyeriss accelerator.}
    \label{fig:energy}
    \vspace{-1.0em}
\end{figure}

Figure~\ref{fig:energy} shows energy breakdowns for the whole MobileNetV1 and various uniform quantization settings. The $x$b denotes the setting for which each layer was independently calculated when $q_a=q_w=q_o=x$. The best mapping for each setting was obtained by random search.
As evident, the energy tied to the memory transfers decreases with decreasing $x$, as it impacts the utilization of memory subsystems that store the respective data. For example, in the case of 4-bit quantization, when compared with the 8-bit setting, the total energy is reduced by more than $32.5\%$, while the energy consumed in memory by $54.5\%$. Moreover, note that for $x \geq 6$, the bit-packing yields no benefit for the 16-bit word size. We can conclude that the proposed Timeloop extension opens new mappings and allows us to exploit them to find more energy-efficient memory utilization.

\subsection{Automated quantization}

The goal of NSGA-II is to find the most suitable trade-offs between CNN accuracy and EDP for one inference. We selected Eyeriss and MobileNetV1 performing ImageNet classification to demonstrate the behavior of the proposed automated quantization algorithm. 

\begin{figure}[ht]
    \vspace{-0.6em}
    \includegraphics[width=\columnwidth]{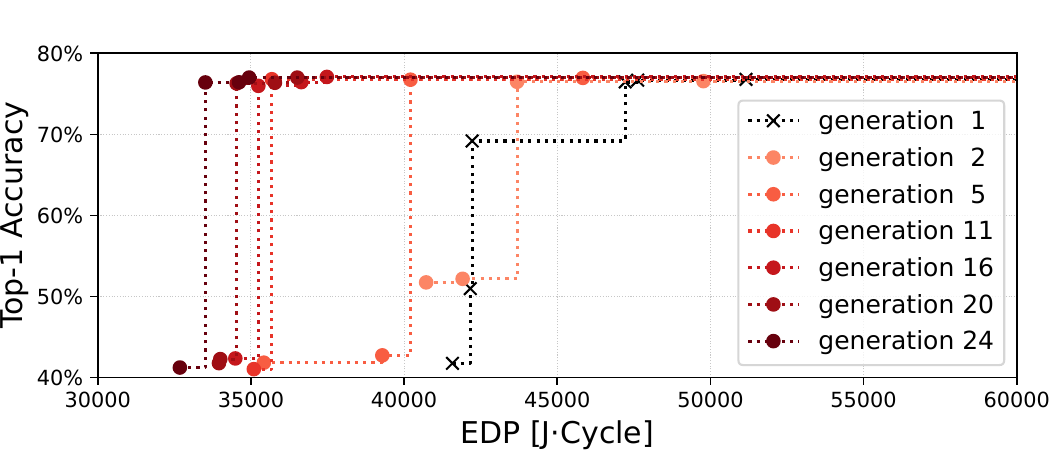}
    \vspace{-1.2em}
    \caption{Pareto fronts showing the best solutions found by NSGA-II across various generations, applied to MobileNetV1.} 
    \label{fig:nsga_generations}
    \vspace{-1.5em}
\end{figure}

On one typical run ($e=10, |Q|=16$), plotted in Fig.~\ref{fig:nsga_generations}, the optimization progress is investigated. The Pareto front is improved significantly compared to the initial uniform quantization. Most of the changes are done before the 11th generation.

As QAT is the most time-consuming part of the optimization process, we have to determine the most suitable parameters of the proposed method to maximize its benefits.

Firstly, we investigate the impact of a pre-trained model on the accuracy achieved after QAT. For the FP32 model, we fine-tuned uniformly quantized configurations for $e = 10$, whereas for the QAT-8 model, we did the same for $e = 5$. The results in Fig.~\ref{fig:final}a show that better accuracies are obtained when QAT-8 model is used, as it is already more accustomed to the effects of quantization. Based on this outcome, we decided to use QAT-8 as the initial model in all of the NSGA-II experiments.

Secondly, we ask what is the best offspring size if the number of evaluations (i.e., the product $|Q| \times$~the number of generations) is constant.
When using the larger offspring population $|Q|$, it turns out that $|Q|=32$ is unsuitable because only 8 generations are generated. There is no significant difference in the resulting Pareto fronts for $|Q|=8$ or $32$, as shown in Fig.~\ref{fig:final}b.  
Please note that the size of the parent population ($|P|$) has only a minimal effect on the performance because these individuals are stored from the previous generations, and their evaluation is already known.

Thirdly, we investigate the best trade-offs between the number of generations and the number of epochs $e$ allowed for training candidate quantized CNNs. Fig.~\ref{fig:final}c shows that better trade-offs between accuracy and EDP are reached when $e$ is higher. However, the search is faster for lower $e$ because the number of generations is 28 and 14 for $e$ = 10 and 20, respectively. 
We prefer a larger number of epochs to achieve better accuracy-EDP trade-off.

\subsection{Final results}

\begin{figure}[ht]
    \centering
    \vspace{-1em}
    \includegraphics[width=\columnwidth]{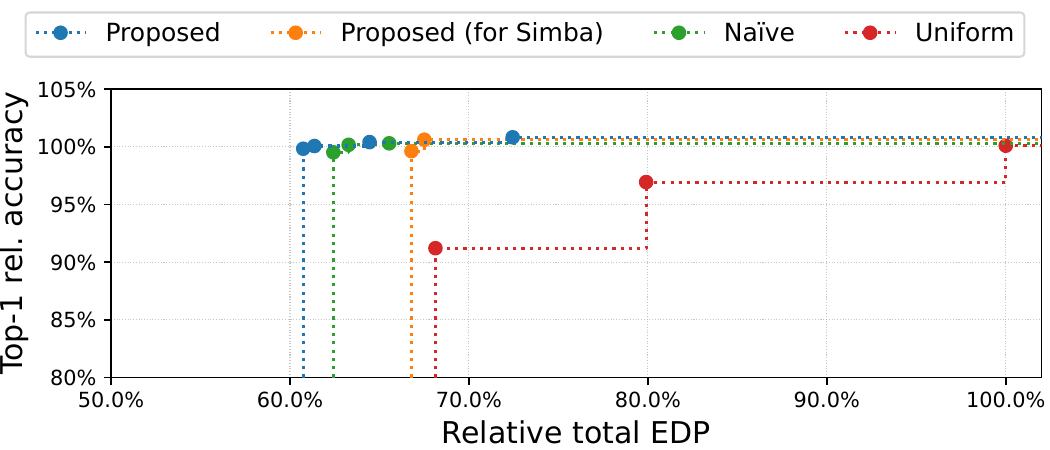}
    \caption{Tradeoff between accuracy and EDP evaluated on Eyeriss accelerator running MobileNetV1 inference. The parameters are relative to the uniform 8-bit implementation.}
    \label{fig:final:comp}
    \vspace{-1em}
\end{figure}
For the final comparisons and based on the previous experiments, we used the following settings of the proposed method: $e=20, |P|=32, |Q|=16 $. 

The proposed method utilizing mixed-precision quantization for the Eyeriss accelerator is compared with the setting employing (a) the uniform quantization (see Uniform in Fig.~\ref{fig:final:comp}), (b) na\"ive quantization (see Na\"ive in Fig.~\ref{fig:final:comp}), and a different accelerator (see Proposed for Simba in Fig.~\ref{fig:final:comp}). Fig.~\ref{fig:final:comp} reveals that if the optimization algorithm exploits the properties of the target Eyeriss accelerator, it can reach better results than for an unseen accelerator (represented by Simba in our case). A na\"ive approach considering only memory aspects reached Pareto front incomparable with the proposed one. Neither the uniform quantization is able to deliver better results than our approach.

Table \ref{tab:final} shows the energy savings in the memory subsystem for two different CNNs and two accelerators in the context of the accuracy drops on the ImageNet subset. The uniform approach represents the SoA solutions that do not explore the quantization of individual layers. It can be seen that this approach has managed to find a small number of relatively good solutions, but worse than when mixed-quantization is considered. The na\"ive approach, which is compared against approaches that do not exploit accelerator properties, delivered good results, but the proposed mixed-quantization method is better. 
The best trade-offs are reported for the Eyeriss accelerator because its memory subsystem has a higher impact on the overall energy.

Our method can also help in designing new hardware accelerators for CNN because it can cheaply estimate the impact of complex quantization schemes on the resulting performance and other parameters without the need to implement the accelerator.

\newcommand{\qchange}[2]{$e_m:$ #1\% $\rightarrow$ $a:$ #2\%}
\renewcommand{\qchange}[2]{$\Delta_e_m$ = #1\% ; $\Delta_a$=#2\%}
\renewcommand{\qchange}[2]{#1\% & #2\%}
\newcommand{\bstart}{\begin{tabular}{@{}cc@{}}}
\newcommand{\bend}{\end{tabular}}

\begin{table}[ht]
    \centering
    \vspace{-0.5em}
    \caption{Reduction in memory energy $\Delta_e{_m}$ and relative accuracy loss $\Delta_{acc}$ achieved by different types of automated quantization algorithms.}
    \label{tab:final}
    \resizebox{\columnwidth}{!}{
    
    \begin{tabular}{c|c|c|c|c}\toprule
        \bf Architecture & \multicolumn{2}{c|}{ \bf{Eyeriss}} & \multicolumn{2}{c}{\bf{Simba}}\\
         \bf Network  & MobileNetV1 & MobileNetV2 & MobileNetV1 & MobileNetV2 \\
         & \bstart $\Delta_e{_m}$ & $\Delta_{acc}$ \bend &  \bstart $\Delta_e{_m}$ & $\Delta_{acc}$ \bend &  \bstart $\Delta_e{_m}$ & $\Delta_{acc}$ \bend &  \bstart $\Delta_e{_m}$ & $\Delta_{acc}$ \bend \\\midrule
         Uniform $^{1)}$ &  \bstart 
            \qchange{-34.9}{-3.0}\\
            \qchange{-54.3}{-8.8}\\
        \bend & \bstart
            \qchange{-33.0}{-0.7}\\
            \qchange{-47.9}{-7.9}\\
        \bend & \bstart 
           \qchange{-34.3}{-3.0}\\
            \qchange{-51.5}{-8.8}\\
        \bend & \bstart
            \qchange{-27.3}{-0.7}\\
            \qchange{-41.0}{-7.9}\\
        \bend\\\midrule
         Na\"ive $^{2)}$ & \bstart
            \qchange{-57.2}{+0.2}\\
            \qchange{-63.4}{-0.5}\\
        \bend & \bstart
            \qchange{-30.5}{+0.9}\\
            \qchange{-41.3}{+0.8}\\
            \qchange{-47.6}{+0.5}\\
            \qchange{-49.2}{+0.4}\\
        \bend & \bstart
            \qchange{-48.7}{+0.3}\\            
            \qchange{-58.0}{-0.5}\\
        \bend  & \bstart 
            \qchange{-25.9}{+0.9}\\
            \qchange{-34.5}{+0.8}\\
            \qchange{-44.2}{+0.7}\\
            \qchange{-49.7}{+0.3}\\
        \bend\\\midrule
         \bf Proposed & 
         \bstart
            \qchange{-45.2}{+0.8}\\
            \qchange{-56.8}{+0.4}\\
            \qchange{-61.9}{+0.1}\\
            \qchange{-63.1}{-0.4}\\
         \bend &   \bstart
            \qchange{-44.2}{+1.3}\\
            \qchange{-46.4}{+0.8}\\
            \qchange{-48.0}{+0.6}\\
            \qchange{-53.6}{+0.5}\\
        \bend & \bstart
            \qchange{-51.7}{+0.4}\\
            \qchange{-53.9}{-0.0}\\
            \qchange{-58.0}{-0.5}\\
            \qchange{-58.6}{-0.5}\\
        \bend  & \bstart
            \qchange{-43.5}{+0.9}\\
            \qchange{-44.2}{+0.8}\\
            \qchange{-47.5}{-0.2}\\
            \qchange{-48.4}{-1.3}\\
        \bend
         
         \\\bottomrule
         \multicolumn{5}{l}{\footnotesize $^{1)}$ Uniform quantization has been used e.g. in \cite{Ristretto,Chen:Eyeriss:2017}.} \\
         \multicolumn{5}{l}{\footnotesize $^{2)}$ A general automated quantization approach without exploring target accelerator parameters e.g., \cite{choi:pact}.} \\
        
    \end{tabular}
    }\vspace{-2em}
\end{table}

\section{Conclusions}
We show that the introduction of mixed-precision quantization of weights and activations in hardware accelerators of CNNs increases the number of available CNN-to-hardware mappings and allows for better trade-offs between energy and CNN errors than common quantization schemes. In contrast to other methods, e.g.~\cite{HAQ:2019}, these trade-offs can be investigated and quantified for general accelerator architectures in software (i.e., in a design phase of a HW accelerator) because the proposed extension of the Timeloop estimation tool has this capability.

{
\noindent\textit{Acknowledgement}  This work was supported by the Czech Science Foundation project 24-10990S and by the Ministry of Education, Youth and Sports of the Czech Republic through the e-INFRA CZ (ID:90254).
}

\bibliographystyle{IEEEtran}
\bibliography{IEEEabrv,date24}

\end{document}